\documentclass[amsmath,amssymb,
 aps]{article}

\usepackage[english]{babel}

\usepackage[letterpaper,top=2cm,bottom=2cm,left=3cm,right=3cm,marginparwidth=1.75cm]{geometry}
\usepackage{cite} 
\usepackage{natbib} 

\usepackage{amsmath}
\usepackage{aas_macros}
\usepackage{graphicx}
\usepackage[colorlinks=true, allcolors=blue]{hyperref}

\title{The fallacies of $\Lambda$CDM falsifications}
\author{Alain Blanchard}

\vspace{-8ex}
\date{}

\begin{document}

\maketitle
\noindent Universit{\'e} de Toulouse, UPS-OMP, IRAP, CNRS, 14 Avenue Edouard Belin, F-31400 Toulouse, France

\begin{abstract} 
In recent years, numerous arguments have emerged suggesting that the $\Lambda$CDM (Lambda Cold Dark Matter) model may be inconsistent with observational data, requiring more or less radical revisions. Notable examples include the ‘Hubble tension’—the discrepancy between early- and late-universe measurements of the Hubble constant—as well as tensions in measurements of cosmic structure growth. These issues have led some to question the validity of the $\Lambda$CDM framework and consider possible modifications or alternative models.

However, upon closer inspection, many of these critiques stem from methodological or interpretive disagreements rather than from clear falsifications in the strict Popperian sense. Karl Popper proposed that scientific theories must be testable and falsifiable; in other words, a theory should be rejected if it fails a specific, reproducible test. Yet, many of the alleged 'inconsistencies' within $\Lambda$CDM, while statistically significant, are not definitive falsifications but rather indicators of areas needing refinement or more complex modeling within the same framework.

Thus, I review the recent claims about $\Lambda$CDM's limitations and analyze why they often reflect individual biases or philosophical preferences, rather than rigorous scientific falsifications. For example, alternative cosmological models such as MOND (Modified Newtonian Dynamics) or models incorporating new physics like quintessence or modified  gravity are sometimes advocated based on theoretical appeal rather than direct evidence from critical tests. In many cases, these arguments for falsifying $\Lambda$CDM reveal more about subjective interpretations of data than about concrete observational contradictions.

\end{abstract}

\section{Introduction} 

The idea of a non-baryonic matter component contributing significantly to the density of the universe dates back to the 1960s : in 1966, Gershtein and  Zel’dovich \citep{PismaZhETF.4.174} noticed that an upper limit to the neutrino mass comes from their possible contribution to the density of the universe. By the 1980s, the possibility of a critical-density universe dominated by neutrinos with masses exceeding several electron volts was explored, largely driven by the work of Yakov Zel’dovich and the claim of the detection of non-zero mass \citep{1980PhLB...94..266L}. It was realized that in such case the  free streaming of neutrinos will smooth out primordial fluctuations on small scales. This hypothesis marked an important shift in cosmological thinking: specifying the nature of dark matter meant that the initial spectrum of density fluctuations was no longer purely determined by unknown initial conditions but instead emerged from the dynamic evolution of perturbations. This led to specific and well-defined features in the final distribution of matter, intrinsically tied to the properties of non-baryonic dark matter.

Neutrinos with masses on the order of a few tens of electron volts (eV) introduced a characteristic suppression scale that was too large to be consistent with the observed distribution of galaxies known at the time. This suppression arose because neutrinos, being relativistic and freely streaming in the early universe, erased density fluctuations below a certain scale. As a result, structure formation was severely inhibited on small scales, preventing the efficient formation of galaxies and galaxy clusters. These fundamental issues ultimately led to the exclusion of the neutrino-dominated scenario as a viable model for dark matter \citep{1983ApJ...274L...1W}. The inability of hot dark matter to support the observed large-scale structure prompted a shift in focus toward alternative dark matter candidates.

By the late 1970s, it was realized that a weakly interacting particle with a mass on the order of a giga-electron volt (GeV) could also contribute significantly to the total density of the universe. Unlike neutrinos, such a particle would become non-relativistic much earlier in cosmic history, allowing density perturbations to survive on small scales and facilitating the hierarchical growth of structures. This idea laid the foundation for the cold dark matter (CDM) paradigm, which would later become a cornerstone of modern cosmology.

In 1982, Jim Peebles \citep{1982ApJ...263L...1P} proposed an alternative scenario based on such  a much more massive dark matter particle, giving rise to what is now known as the cold dark matter (CDM) paradigm. He noted that the galaxy correlation function roughly agreed with observations and that this model predicted a significantly lower amplitude for cosmic microwave background (CMB) fluctuations than in a purely baryonic universe. Around the same time, the concept of cosmic inflation was gaining traction. Inflation naturally led to a specific initial fluctuation spectrum—now known as the Harrison-Zel’dovich spectrum—providing a logical foundation for a cosmological model with very few free parameters. In this framework, the spectrum of density perturbations was determined by just a handful of cosmological parameters and the overall amplitude of fluctuations.

Furthermore, the evolution of perturbations in the early universe produced a transfer function with multiple characteristic scales. The precise values of these scales depended on the details of the cosmological model, particularly on key parameters such as the baryonic matter density, the non-baryonic dark matter density, and the Hubble constant. Initially, this cold dark matter model was developed under the assumption of a universe with critical density. However, as observational data on small-scale CMB fluctuations started to emerge, it became evident that the universe was nearly flat, yet no dynamical measurement indicated a total matter density close to the critical value. This discrepancy led to the proposal of introducing a cosmological constant ($\Lambda$) to the model, which eventually evolved into what is now known as the $\Lambda$CDM model.

One of the most significant strengths of this framework is its predictive power. Given that the matter spectrum is approximately fixed, it allows for precise predictions of both the CMB fluctuation spectrum and the large-scale matter distribution (and thus the distribution of galaxies, up to a bias parameter) with very few free parameters. This predictive capability was put to the test when full-sky measurements of CMB fluctuations became available with the launch of the Wilkinson Microwave Anisotropy Probe (WMAP) in 2003: once normalized to the CMB,  the $\Lambda$CDM model predicted the entire linear  matter power spectrum (or equivalently, the correlation function). Strikingly, when data from the Sloan Digital Sky Survey (SDSS) became available in 2005, the observed large-scale structure exhibited an almost perfect match with the predictions of the $\Lambda$CDM model—an extraordinary success for what is now considered the “standard model” of cosmology.

Since then, the robustness of this agreement has been further reinforced by increasingly precise CMB measurements from WMAP and later from the Planck satellite, as well as by improved large-scale structure data from SDSS and other galaxy surveys. Today, no alternative scenario offers a comparably successful description of cosmic structures on large scales. The strong predictive power of the standard $\Lambda$CDM model continues to enable rigorous and fruitful comparisons with observational data. This makes it possible to envisage its falsification, but also makes it the dominant framework for our understanding of the large-scale evolution of the universe.

\section{What Makes a Scientific Theory Successful?}

Scientific inquiry does not seek to unveil an absolute reality, let alone an ultimate truth. Instead, it constructs models that describe and predict observable phenomena, always remaining open to revision in light of new evidence. A defining feature of modern science is its commitment to the Popperian framework: a scientific theory must not only explain known facts but also predict new and unexpected phenomena. Crucially, these predictions should be as distinct as possible from previously established knowledge, ensuring that the theory is testable and falsifiable.

A historical example of this principle is Newton’s theory of gravity, which reached its pinnacle when it successfully predicted the existence of Neptune. The discovery was based on unexplained anomalies in Uranus’ trajectory, leading astronomers to infer the presence of an unseen planet. This validation of Newtonian mechanics remains a landmark in the philosophy of science, demonstrating the predictive power that distinguishes scientific theories from mere descriptions. Of course, this success was even more striking than Newton’s ability to account for planetary orbits, as it showcased the theory’s capacity to anticipate phenomena beyond those it was originally formulated to explain.

Similarly, it is crucial to examine how the $\Lambda$CDM model has undergone the Popperian falsification test—and passed with flying colors. However, the Popperian sequence of prediction, testing, and potential falsification should not be interpreted naively as a strict chronological process. What must prevail in the scientific process is the logical structure of the sequence of events. An unexpected discovery that is later shown to be deducible from a pre-existing theory or model constitutes a successful Popperian test. The fact that the phenomenon was not initially predicted, or that it was not announced as a test of the theory, but that it can be deduced, even retrospectively, from the theoretical framework reinforces confidence in the model, because it demonstrates its predictive power. This aspect is discussed in a little more detail by \citep{2025arXiv250106095F}. 
This process exemplifies the self-correcting nature of science: theories are not only expected to explain known facts but also to withstand scrutiny when new, unforeseen phenomena emerge,  or even show that this new observation follows naturally from the theory. An emblematic example of this situation is Hubble's discovery in 1929: the proportionality between the redshift of distant sources, as galaxies were, and their distance had been noticed and recorded in cosmological models derived from general relativity, though it had never been explicitly stated as a prediction. Nevertheless, Hubble's observations were regarded at the time—and still are today—by the scientific community as the first validation of relativistic cosmological models \footnote{ There is a certain irony in this event in two ways: Einstein modified his initial theory because he was focused on a static cosmological solution; the equations of an expanding universe can be deduced from Newtonian reasoning, they could therefore have been deduced before the advent of general relativity.}

In contrast, if a new fact arises and the theory is modified after the fact by introducing additional ad hoc elements solely to accommodate it, the epistemic value of the test is considerably weakened—or even nullified altogether. A theory that is excessively flexible, capable of adapting to any new data simply by incorporating arbitrary adjustments, risks becoming unfalsifiable, thereby stepping outside the bounds of empirical science. However, this does not mean that theories cannot evolve. The Popperian test remains meaningful if the revised version of the theory is not merely retroactive accommodation but instead introduces new, independently testable predictions. In such cases, the modifications are not ad hoc but part of a refined theoretical framework that remains falsifiable. The key is that any refinement should lead to novel, discriminating, and risky predictions—ones that could, in principle, prove the new formulation wrong. Only under these conditions does scientific progress remain genuinely empirical rather than purely adaptive.

That said, although Popperian falsifiability is an important methodological ideal, science operates more flexibly in practice. Many theories are accepted because they are coherent, explanatory, and drive progress, even if they are not immediately testable. The robustness of a theory often relies on an accumulation of converging evidence rather than a single falsification test. For instance, string theory is mathematically elegant and consistent with known physics, but it still lacks testable predictions that can be verified by current experiments. Likewise, the theory of evolution is not easily testable through a strict Popperian sequence, yet it is strongly supported by an accumulation of independent observations.

Furthermore, the history of science teaches us that even the most successful theories have their limits. Einstein’s general relativity ultimately refined Newton’s framework by incorporating the curvature of spacetime, providing deeper insights into gravity. This iterative process underscores the self-correcting nature of science: theories evolve, but they are never final. The scientific endeavor remains an ongoing process of refinement, where theories must continuously prove their explanatory and predictive power to retain their status as reliable models of reality.

A final remark concerns the nature of the facts used in the Popperian criterion. An interesting historical example is the solar neutrino deficit. For the record, models of the Sun’s interior predicted a flux greater than that measured by early neutrino detectors, such as the Homestake experiment. However, doubts arose regarding the validity of these stellar interior models, given their reliance on complex and sometimes untestable hypotheses. This situation is not uncommon in astrophysics: predictions often depend on intricate modeling, which can weaken their robustness—not necessarily due to inaccuracies, but because they are grounded in assumptions that may themselves lack direct empirical validation. The key takeaway is that the strength of a Popperian test involves a degree of subjectivity: it must be regarded as robust by the scientific community as a whole, which may not be the case for models of the Sun's interior.
The resolution of the solar neutrino problem came not from astrophysics but from particle physics, with the groundbreaking discovery of neutrino oscillations—a phenomenon demonstrating that neutrinos can change “flavor” as they travel through space. This insight not only solved the deficit but also provided experimental confirmation that neutrinos have a small but non-zero mass. Subsequent experiments, such as those conducted at Super-Kamiokande and the Sudbury Neutrino Observatory (SNO), validated both the phenomenon of oscillations and the accuracy of solar interior models, marking a milestone in our understanding of both particle physics and stellar processes.
In this article, I aim to examine how the $\Lambda$CDM model has been tested against observational data and how no alternative provides the same level of Popperian corroboration. 

\section{Successes of the $\Lambda$CDM model}

There is no precise date that marks the birth of the $\Lambda$CDM model. As mentioned earlier, the $\Lambda$CDM model itself is difficult to date. The role of the cosmological constant has been studied since Einstein first introduced it, but for a long time, the lack of observational evidence made it an exotic concept in cosmology. J. Peebles later revisited this idea to explain a flat universe without requiring the matter density to be equal to the critical density \citep{1984ApJ...284..439P}. He then developed what is now known as the quintessence scenario, in which the cosmological constant is replaced by a scalar field whose behavior, depending on its potential, can mimic that of a cosmological constant \citep{1988PhRvD..37.3406R}. The introduction to this article explains that one of the motivations is to achieve a flat universe, which requires a critical density, without invoking non-baryonic dark matter. 

By the late 1990s, all the key ingredients of the $\Lambda$CDM model were in place: baryonic matter, cold non-baryonic matter, and dark energy. The theoretical framework for computing the transfer function had been established and refined since the work of Peebles and Yu \citep{1970ApJ...162..815P}. Once these components were specified, the full calculation of the fluctuation spectrum in the cosmic microwave background (CMB) and the matter distribution spectrum became well-defined.

A crucial element in this framework is the initial spectrum of fluctuations. From the 1970s onward, for simplicity, it was generally assumed that this spectrum was Gaussian and followed a power law. Whether the fluctuations were adiabatic or not, however, remained a subject of debate. The emergence of inflation theory provided a theoretical foundation for these assumptions. But it should be kept in mind that by the 1980s, the essential ingredients now incorporated into Boltzmann codes such as CAMB and CLASS were already in place.

It is remarkable how observations of the cosmic microwave background (CMB) and the large-scale distribution of galaxies have consistently supported the $\Lambda$CDM model over the past four decades. For the record, when Peebles wrote his article on CDM \citep{1982ApJ...263L...1P}, no fluctuations had been measured in the CMB, and there was no three-dimensional catalogue of galaxy distribution
 no three-dimensional catalogues were available to study the distribution of galaxies (only projected distribution was well studied).
Some alternative perspectives, such as the hot dark matter model and the cosmic string scenario, were quickly abandoned: hot dark matter due to galaxy distribution data ruling out the massive neutrino scenario. Models of structure formation based on topological defects, such as textures, were gradually ruled out as increasingly precise observations of cosmic background fluctuations became available. 

The measurement of the large-scale galaxy angular correlation function was the first major challenge to the standard CDM model (i.e., the critical density model) \citep{1990MNRAS.242P..43M}: Indeed, the shape of the angular correlation function obtained from the APM was inconsistent with the expectations of a CDM model with a critical density. However, it was observed that a CDM-type model with a nonzero cosmological constant provides a good fit to the data \citep{1990Natur.348..705E}. We could consider this observation as marking the operational birth of the $\Lambda$CDM model. 

\subsection{The fluctuations of the cosmic microwave background}

The first detection of CMB fluctuations by COBE took place only a few years after the measurement of the APM correlation function.
The shape and amplitude of these fluctuations, as well as their Gaussian or non-Gaussian nature, had the potential to strongly challenge—or even outright invalidate—the $\Lambda$CDM scenario.
In retrospect, it is interesting to note that, although the signal detected by COBE matches precisely what is expected in the $\Lambda$CDM model, this agreement was not particularly emphasized at the time.
Yet, this is one of the most astonishing results of the theory: COBE's observations reveal the amplitude of matter fluctuations on scales 100 to 1000 times larger than those accessible to robust measurements from galaxy catalogues at the time, and their amplitude agrees to within a factor of order unity with theoretical predictions. 

The fundamental role of CMB data in validating the $\Lambda$CDM model cannot be overstated here: the calculation of CMB fluctuations is highly robust, as it is performed within the linear perturbation regime. This ensures that no ad hoc or overly simplifying assumptions have been introduced. Moreover, the various existing codes—some of them independent and competing—agree to an extremely high degree of accuracy. The measurement of these fluctuations is also highly reliable: Galactic foregrounds, while present, are not dominant and can be efficiently subtracted due to their frequency-dependent characteristics\footnote{
It remains an interesting question whether over-simplistic assumptions have been made about the properties of astrophysical foregrounds, such as dust, which would lead to a (slight) bias in the cosmological parameters.}. Finally, after more than thirty years of improvements, Planck’s results are fully consistent with those of WMAP, as well as with the most recent ground-based experiments, such as ACT and the South Pole Telescope. 

After the first version of this paper was submitted, the ACT team published the DR6 data and constraints. This is an excellent Popper test: once the model is fitted to previous CMB data, it predicts very accurately the properties of the power spectrum of the fluctuations $C_l$ . ACT provides new measurements at small scales, increasing by nearly of factor of two the scales at which accurate measurements are available. The data are fully consistent with $\Lambda$CDM \citep{2025arXiv250314452L}, constraining severely current alternatives that modified the small scale behavior of the $C_l$ \citep{2025arXiv250314454C}.

\subsection{The large scale structure of the universe}

The measurement of the large-scale structure of the universe in a quantitative way is primarily performed through the correlation function of galaxies. As we have already mentioned,  Peebles \citep{1982ApJ...263L...1P} pointed out that the shape of the power spectrum was in {\em qualitative} agreement with what was know at that time. Forty years on, observations of the distribution of galaxies on linear scales, particularly at the BAO scale, remain in perfect agreement with theoretical predictions, even though they could have invalidated the model, as was the case for massive neutrinos. The DESI data provides the latest tight constraints \citep{2024arXiv241112022D}. 

\subsection{The global picture} 

The fact that the $\Lambda$CDM model is consistent with both the CMB and the large-scale structure (LSS) distribution of galaxies, while measurements reach an accuracy on the order of one percent, is a remarkable achievement of modern science that should not be underestimated. The ability to determine cosmological parameters independently from these two datasets, yielding consistent values with an accuracy of a few percent, is an impressive test of consistency.

\section{Criticism of the $\Lambda$CDM model}

Many articles deal critically with the Standard Model. But the criticisms are of different kinds. There's also a low-level skepticism among some physicists, stemming from the fact that in cosmology, there are virtually no laboratory experiments to back up the model, to the constraints of, say, atomic or particle physics. The impression is therefore that the assertions are based on a small number of observations that provide only tenuous support for the representations that are elaborated. 

\subsection{Early Universe and initial conditions}

In order to make fruitful comparison of $\Lambda$CDM model with observations it is necessary to assume some initial power spectrum, in practice a power law:
\begin{equation}
P(k) \propto k^n
\end{equation}
with a free index $n$ close to 1. Furthermore, initial fluctuations are assumed to be adiabatic and Gaussian. These ingredients are easily produced by in inflation models. 
By way of an intellectual shortcut, it is often claimed that the $\Lambda$CDM model relies on inflation—considered by some to be a speculative theory with limited observational validation, and some even argue that it is contradicted by the data \citep{2014PhLB..736..142I}.
$\Lambda$CDM would therefore suffer from the same weakness \citep{2023eppg.confE.231K}. However, key ingredients of the model were already in use before the formulation of inflation, such as adiabatic fluctuations with a power-law spectrum and a Gaussian distribution. Whether or not inflation is correct is therefore irrelevant to the $\Lambda$CDM framework, as the defining characteristic of Cold Dark Matter (CDM) lies in the transfer function—that is, in how fluctuations evolve over time from their initial state. The introduction of a cosmological constant became necessary when observations of the cosmic microwave background indicated a flat universe, even before the supernova Hubble diagram provided direct evidence of cosmic acceleration.

\subsection{The large scale anomalies}

\subsubsection{(in)Homogeneity on large scale}
The standard cosmological model is based on Einstein's cosmological principle, which states that the universe is homogeneous on large scales, allowing the use of the Robertson-Walker metric. However, this remains a qualitative statement of a question that must ultimately be addressed in quantitative terms. To achieve this, it is relevant to examine the amplitude of metric perturbations. As long as these perturbations remain very small, they can be treated to linear order, with higher-order corrections included if observational precision warrants it. In the solar system, for example, the Sun and planets introduce only weak perturbations, and relativistic effects can be accounted for by considering higher-order terms. The Sun represents a perturbation of the metric of the order of $10^{-5}$, even though the density contrast is enormous. On the scale of a galaxy, the perturbation is of the same order. On larger scales, galaxy catalogs can be used to assess density contrast. It has long been known that : 
\begin{equation}
\frac{\delta \rho}{\rho}(10h^{-1}\mathrm{Mpc})\sim  1 
\end{equation}

At larger scales, the density field is expected to be linear (i.e., fluctuations are expected to be less than one). However, visual inspection reveals structures that appear much larger, giving the impression that very large inhomogeneities might exist. These types of features are sometimes perceived as challenges to the standard picture \citep{1986ApJ...302L...1D}, but they do not hold up to further quantitative analysis. This argument has been reiterated recently  to invalidate the $\Lambda$CDM \citep{2024arXiv240914894L}, but comparisons with numerical simulations remain ultimately inconclusive \citep{2025arXiv250203515S}.

\subsubsection{Dipole}

Several surveys of distant sources have identified a dipole anisotropy in their distribution \citep{2023CQGra..40i4001A}. If this is interpreted as a kinematic dipole, it exceeds that of the CMB. However, to my knowledge this argument never consider the possibility that the dipoles in sources is just due to the randomness of the clustered process. Given the above fluctuation, assuming that the universe is filled with random spheres of $10h^{-1}\mathrm{Mpc}$ one would expected fluctuations of the order of few $10^{-3}$, comparable or larger than the CMB dipole. A second problem that arises is that, in order to be confident that the dipole is real, one must control the selection function of objects across the sky with an accuracy of at least $10^{-3}$,  which is highly challenging.

\subsection{The $S_8$ tension}

The amplitude of matter fluctuations  in the present day universe is traditionally measured by $\sigma_8$.  The abundance of clusters and the weak lensing measurements are two traditional ways to evaluate its amplitude and it is cosmology rephrased in term of $S_8$:
\begin{equation}
S_8 = \sigma_8\left(\frac{\Omega_m}{0.3}\right)^{1/2}\,,
\end{equation}
which  is less sensitive to $\Omega_m$.   Lensing measurements have long yielded lower values than those inferred from the CMB in the $\Lambda$CDM framework \citep{2022A&A...665A..56L}. The most emblematic measure is the one obtained from Kids $3\times2$pt \citep{2021A&A...646A.140H}:
\begin{equation}
S_8 = 0.766 \pm 0.02
\end{equation}
Indeed, the amplitude of matter fluctuations at redshift $\sim 1100$ can be inferred from CMB fluctuations. Planck provides the most accurate value, consistent with other CMB measurements. In order to compare this  to present day values, one must extrapolate it to the $z = 0$.
This is straightforward in the  in the $\Lambda$CDM model. However, it is important to note that the inferred  $S_8$ value is model dependent. The tension with Kids was 3$\sigma$.  

The situation has recently changed in recent years. The Planck latest value obtained from the PR4  release \citep{2024A&A...682A..37T} is slightly lower than the 2018 value: 
\begin{equation}
S_8 = 0.819 \pm 0.014
\end{equation}
lowering the tension down to $2.2 \sigma$. Moreover,  the DES and KiDS teams have reached a joint conclusion, finding a value regarded as consistent with Planck \citep{2023OJAp....6E..36D}:
\begin{equation}
S_8 = 0.790^{+0.018}_{0.014}
\end{equation}
the two $S_8$ now differs by only 1.3 $\sigma$.
In fact, several recent measurements based on low-redshift data consistently yield a value compatible with that inferred from the CMB. In \citep{2024OJAp....7E..32B} using a combination of different local measurements we obtained:
\begin{equation}
S_8 = 0.801 \pm 0.012
\end{equation}
I think it's fair to say that the tension in $S_8$ appears to no longer exist. Ironically, the eROSITA collaboration's analysis of X-ray clusters abundance suggests a $S_8$  higher value than that derived from the CMB, potentially indicating a new tension with an $S_8$ value higher than what is inferred from the CMB...

\subsection{"Crises" on non-linear scales}

On small scales (i.e., non-linear scales), numerous claims suggest that the predictions of the 
$\Lambda$CDM model are inconsistent with observations. These include the core-cusp problem, the Too Big to Fail problem, the missing satellites problem, the satellite planes problem, and the baryonic Tully-Fisher relation. An extensive body of literature has explored these issues (see, for instance, the comprehensive review \citep{2022NewAR..9501659P}).

Assessing the significance of these problems is challenging for several reasons. The first issue is the reliance on a posteriori statistics: a striking pattern is identified and then judged to be highly improbable. For example, in the case of the satellite planes problem, several satellites of the Milky Way and Andromeda were observed to be located within a thin plane \citep{2013Natur.493...62I}. However, such configurations were later found to be not so uncommon in numerical simulations and consistent with expectations in 
$\Lambda$CDM \citep{2023NatAs...7..481S,2024MNRAS.535.3775U}.
The second problem is that numerical simulations are required to address the issue, as in the previous example. Indeed, the core-cusp problem, the Too Big to Fail problem, and the missing satellites problem are based on non-linear features that can only be addressed through numerical simulations. Because we are on non-linear scales, the interpretation of data may be misleading \citep{2004ApJ...617.1059R}. A final difficulty is that, most of the time, the 'predictions' of the $\Lambda$CDM  model are derived from pure dark matter simulations. It is well known that hierarchical model of structure formation are subjected to the "overcooling" problem: that is simple estimation of the amount of primordial gas that should ended in a "cold form" (i.e. condensed in structures below $10^4$ K) is far in excess to the what is seen in the universe \citep{1992A&A...264..365B}. This is a strong indication that feedback mechanism were strong in the past universe, and the role of baryonic physics could be critical for the interpretation of some data \citep{2023A&A...678A.109A}. This essentially means that  small scale problems should be discussed in a $\Lambda$CDM $+$ baryonic physics and Conclusions drawn from these scales without considering the diversity of possibilities induced by baryonic physics are open to question and cannot, on their own, seriously disqualify the 
$\Lambda$CDM model.

\subsection{The Hubble constant tension}

The  Hubble tension has now become widely recognized and extensively discussed within the scientific community. This discrepancy, which arises from the different values of the Hubble constant  inferred from early- and late-Universe measurements, remains one of the most significant challenges in contemporary cosmology. While  probes based on early-Universe physics, the cosmic microwave background (CMB) and the baryonic  acoustic oscillations (BAO) scale  predict a lower value of around 67 km/s/Mpc, local measurements based on the distance ladder method, including Cepheids and Type Ia supernovae, yield a significantly higher value, around 73 km/s/Mpc. Readers seeking a more comprehensive overview of the Hubble tension, its implications, and the various proposed resolutions can refer to the numerous dedicated reviews and research articles on the subject \citep{2017NatAs...1E.169F,2019NatAs...3..891V}. As stated in \citep{2017NatAs...1E.169F}, the resolution of the tension is either in new physics or in as-yet unrecognized uncertainties. 

There have been numerous efforts to identify possible extensions to the standard $\Lambda$CDM model in an attempt to resolve the Hubble tension. In contrast, comparatively less work has focused on investigating potential systematic biases in measurements.

Most theoretical approaches to addressing the Hubble tension involve extensions of the $\Lambda$CDM framework. This is largely because 
this framework is sufficiently well-defined to enable precise and testable calculations that can be directly compared with observations. Moreover, the accuracy of these observational constraints, notably on the Hubble constant, has now reached the percent level, making it possible to rigorously assess the viability of any proposed modifications to the model. There is a long list of such different models that are proposed to  reduce the Hubble tension \citep{2021CQGra..38o3001D}. A statistical comparison of their merits has led us to rank them in order of success \citep{2022PhR...984....1S}. 

The Hubble tension is undoubtedly the issue attracting the most attention from the community. If confirmed, it would necessitate a revision of the $\Lambda$CDM model. Conversely, if the locally measured Hubble constant ultimately aligns with the value inferred from CMB observations, the $\Lambda$CDM model would receive a strong Popperian corroboration.

The key question here is what conclusions can legitimately be drawn. While it seems widely accepted that the Hubble tension is real, I would argue that it is still too early for definitive conclusions. The measurement of the Hubble constant has undergone numerous revisions since the initial work in 1929, often due to previously unidentified systematic effects. Despite the remarkable efforts of the SH0ES team \citep{2022ApJ...934L...7R}, the possibility of such systematics influencing current measurements cannot be ruled out. The competing team from the Chicago-Carnegie Hubble Program (CCHP) concluded that their measured value of the Hubble constant was consistent with the $\Lambda$CDM model \citep{2024arXiv240806153F}. 

Indeed, unidentified systematics remain the key issue. For instance, a  recent study on the determination of Hubble constant 
  using time delays in gravitational lenses suggests that previous analyses were biased. The new central value obtained is consistent with the $\Lambda$CDM model, but the uncertainties remain too large to resolve the debate \citep{2024arXiv241016171L}. 

  One reason for skepticism toward alternative theories is that they generally introduce additional parameters, which broaden the range of admissible values for the Hubble constant. However, when the SH0ES measurement is not included in the analysis, the central value typically remains close to that obtained from Planck alone. In a recent work we have followed a statistical approach: we compared the merits of the Cepheid calibration bias hypothesis, which is applied to several probes, with those of alternative models. The statistical gain is comparable to that of the best models available \citep{2024OJAp....7E..32B}. The interpretation is that a bias in the SH0ES distance scale due to a miscalibration of Cepheids is as likely as the best alternative models. Recently, \cite{2024PhRvD.110l3518P} pointed out that distance ladder measurements systematically differ from methods that do not rely on them, a conclusion in the same vein.

  \subsection{The matter density ($\omega_m$) tension}

The tension in a cosmological model with data can obviously be assessed using different observables or a combination of them. A common way to express the Hubble tension, for instance, is to compute the following quantity (the Gaussian Tension):
\begin{equation}
\frac{\tilde{H}_0-H_0)_{\rm SH0ES}}{\sigma^2+\sigma_{\rm SH0ES}^2}
\end{equation}
This quantity expresses, in terms of standard deviation, the disagreement between the best estimate 
$\tilde{H}_0$
  obtained from a given data set and the value measured by the SH0ES team, with 
$\sigma$ and $\sigma_{\rm SH0ES}$
  representing the standard deviations for the model and the SH0ES data, respectively. Using the latest SH0ES measurement \citep{2022ApJ...934L...7R} and the combined Planck + DESI data \citep{2024arXiv241112022D}, this results in a 4.56 $\sigma$
tension. (This can be converted into a probability, assuming the statistics follow a Gaussian distribution.)

However, under the assumptions of standard physics in the early universe, certain quantities can be derived from the CMB independently of the late-time evolution of the Universe. Among these quantities is the density parameter $\omega_m$ \citep{2023PhRvD.107j3505L}:
\begin{equation}
\omega_m = 0.14143 \pm 0.0013
\label{eq:LL}
\end{equation}
as we have measurement of $\Omega_m$ in the present day universe, we can compare values of $\omega_m$ for a given  value of the Hubble constant \citep{2024OJAp....7E..32B}.  Combining with DESI $\Omega_m$  we have:
\begin{equation}
\omega_m = 0.17 \pm 0.005
\end{equation}
in a 4.56 $\sigma$
tension with Eq.(\ref{eq:LL}).

\subsection{Nature of  dark matter}

The $\Lambda$CDM model has two major implications for fundamental physics: the existence of a dark matter component and the existence of a source of the acceleration of the expansion of the universe. While the nature of dark energy remains an open question for which only astronomical observations seem likely to provide a better understanding, the status of dark matter is somewhat different. Many experts in the field favor the idea that dark matter is a hitherto unknown massive particle, a WIMP, as originally proposed by J. Peebles \citep{1982ApJ...263L...1P}. This prospect has been and remains a strong driving force for particle physics research. However, it must be remembered that this is a possibility for which there is no evidence. The non-detection of dark matter particles over the last fifty years \citep{2024Symm...16..201M,2024PhRvD.109k2010A} cannot be considered as a falsification of the model: the nature of dark matter is not an explicit ingredient of the model. It could therefore be a question of particles that escape all possibility of detection \citep{2019JCAP...04..005F} or of primordial black holes \citep{2025arXiv250215279C} or generalized dark matter \citep{1998ApJ...506..485H} or some other possibility yet to be discovered...

\subsection{Dynamical dark energy?}

There is now some evidence for the existence of dynamic dark energy \citep{2023arXiv231112098R}, a result that has been confirmed by the analysis of DESI's BAO signal \citep{2025JCAP...02..021A}. If confirmed, this would imply a falsification of the $\Lambda$CDM. 
Dynamic dark energy would be a significant discovery, but rather than fundamentally challenging the structure of the model—as the $H_0$ tension does—it would refine one of its least understood components, namely the source of the accelerated expansion.

\section{Conclusions}

The $\Lambda$CDM model has demonstrated remarkable success in explaining a wide range of cosmological observations, from the cosmic microwave background to large-scale structure formation. Its predictive power, minimal number of free parameters, and internal consistency make it the dominant framework in modern cosmology. However, challenges remain, particularly at small scales, where certain observational features appear difficult to reconcile with theoretical expectations, and in the context of global tensions such as the discrepancy in Hubble constant measurements.

While these issues have sparked debates on potential modifications to 
$\Lambda$CDM or alternative models, they do not yet constitute definitive falsifications in the Popperian sense. Many of the reported tensions could stem from methodological choices, observational systematics, or an incomplete understanding of baryonic physics, which remains a complex and evolving field. Before interpreting them as fundamental inconsistencies of the model, these aspects must be thoroughly examined to assess whether they reflect genuine discrepancies or limitations in current analyses. A deeper understanding of astrophysical processes and improvements in observational techniques will be crucial in clarifying the significance of these tensions.

Among the various challenges, the Hubble constant tension stands out as the most serious. The persistence of a discrepancy between early- and late-Universe measurements has led to speculation about new physics beyond $\Lambda$CDM. However, nearly a century of revisions to Hubble constant measurements has shown that systematic uncertainties have often played a key role in shaping the inferred values. Given this history, resolving the current discrepancy definitively requires an independent, alternative measurement that is not directly tied to traditional distance ladder. Promising avenues, such as gravitational wave standard sirens, offer the potential for such a test \citep{2023PhRvD.108d3543M}, assuming that unforeseen systematics do not arise. A robust, unbiased determination of $H_0$ from these new methods would provide a decisive assessment of whether the tension signals a genuine failure of $\Lambda$CDM or remains within the realm of observational uncertainties.

\bibliographystyle{mnras}
\bibliography{biblio}

\end{document}